# Flexible metro network architecture based on wavelength blockers and coherent transmission

*Annika Dochhan[1], Robert Emmerich[2], Pablo Wilke Berenguer[2], Colja Schubert[2], Johannes K. Fischer[2], Michael H. Eiselt[1], Jörg-Peter Elbers[1]*

[1]*ADVA Optical Networking SE, Märzenquelle 1-3, 98617 Meiningen, Germany*
[2]*Fraunhofer Institute for Telecommunications Heinrich Hertz Institute, Einsteinufer 37, 10587 Berlin, Germany*
adochhan@advaoptical.com

## Abstract

A flexible semi-filterless architecture for metro DWDM networks with ring- or horseshoe topologies is presented. ROADM nodes are based on wavelength blockers. Coherent transmission enables filterless channel selection. Critical parameters are determined and a 4-span, 160-km horseshoe with 100Gbit/s DP-QPSK and 200Gbit/s DP-16QAM is demonstrated.

## 1. Introduction

With the introduction of 5G in the near future, new challenges will be imposed on optical networks in terms of cost-effectiveness, energy efficiency, latency, agility and programmability. The EU H2020 5G-PPP project METRO-HAUL [1] aims at developing the metro-interface between 5G access and multi-Tbit/s elastic core networks, including the optical layer as well as compute nodes with the required software defined networking (SDN) and network function virtualization (NFV). Metro networks typically show non-meshed ring- or horseshoe (open ring) topologies, with link distances depending on the geo-type, e.g. urban or rural. A typical metro horseshoe topology is shown in Fig. 1. Metro core edge nodes (MCEN) are connected to the core network, while access metro edge nodes (AMEN) are connected in a line to form the horseshoe itself. The AMENs include edge data centres (DC) to enable time critical computations close to the end-user, while less latency-sensitive and larger storage requiring functions can be placed in regional or core DC.

While filterless network architectures for core networks with coherent transponders have been proposed already some years ago [2-3], data traffic and flexibility requirements expected from new edge computing nodes now justify the use of coherent technology also in the metro area. Thus, cost-effective filterless nodes only based on splitters have attracted attention recently [4]. Besides the pure splitter-based drop and waste architecture, also semi-filterless solutions have been proposed [5-6], either using filters in some nodes to enable wavelength re-use or to include low-cost filters into the drop-path of the node. This enhances the dynamic range of the received power, which needs to stay between a total allowed maximum for the coherent receiver (Coh Rx) and a minimum power per channel for a certain target performance. In this paper, we focus on the semi-filterless approach using wavelength blockers (WB) inside the nodes. While WB-based reconfigurable optical ring nodes architectures are not new [7], they have recently regained attention in conjunction with coherent transponder technology. The WB devices considered here are based on liquid-crystal-on-silicon (LCoS) technology, similar to [8]. The LCoS chip is divided into four sections, where two enable independent wavelength blocking for two directions and the other two are connected to photo diodes for channel-wise power measurements by a sweeping filter and subsequent power equalization. We show the implications and advantages of this semi-filterless network and demonstrate it in a 90-channel, 4-span, 160-km horseshoe network with DP-QPSK and DP-16QAM, i.e. 100 Gbit/s and 200 Gbit/s channels, respectively.

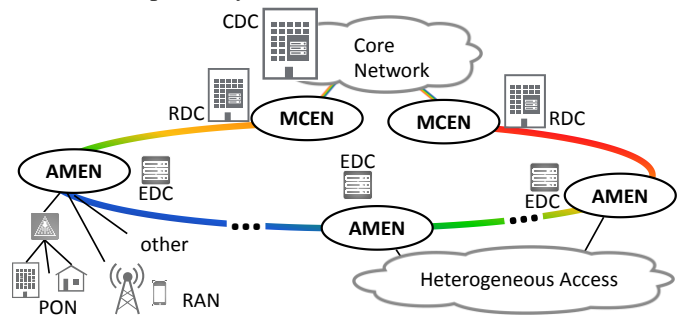

Fig. 1. Metro horseshoe topology. (CDC: central data centre, RDC: regional data centre, EDC: edge data centre, MCEN: metro core edge node, AMEN: access metro edge node).

## 2. Filterless and semi-filterless networks

### 2.1. Structure of the network

Fig. 2 shows the structure of the network. At the MCEN terminal node, $N$ channels (e.g. 96 50-GHz spaced channels in the C-band) are coupled together with couplers, showing ~ 20 dB loss. The coupler is followed by an Erbium doped fibre amplifier (EDFA) as booster and the first fibre span. After the span, an additional EDFA is used as a pre-amplifier for the dropped channels and as a booster for the continuing channels. A further EDFA could be added for very long spans but is not necessary for most urban metro distances. A fraction $R_{drop}$ = 20% of the signal power is coupled out at the drop stage, while 80 % continues to the wavelength blocker. This is the splitting ratio for the devices used in the experiments and should be optimized according to the requirements of the considered network. The 20 % output is followed by a 1x$K$ splitter, to drop $K$ channels which are detected by $K$ coherent receivers. These channels are blocked in the express path by the WB. In addition to the blocking functionality, the power of each remaining channel is measured and equalized. This enables the use of cheaper EDFAs without flattening filters, like in [9]. The $K$ free wavelength slots can be filled with new channels, which are combined by a $K$x1 coupler and inserted into the spectrum by another 20/80 coupler. The resulting signal is transmitted over the next span. At the end of the horseshoe, i. e. at the second MCEN node, the remaining channels are split and fed to coherent receivers.

*2.2. Design parameters and implications*

The limitations in the drop path are determined by the dynamic range of the coherent receivers, i.e. the range from a minimum power $P_{min,Rx}$ per channel (at low optical signal-to-noise ratio, OSNR) that is required to detect an error-free signal to a maximum power $P_{max,Rx}$ accepted by a receiver (Rx), which results from the total power of all channels. Since the output power $P_{amp}$ of cost-effective conventional EDFAs is limited (typically to ~ 20 dBm), the drop ratio has to be $R_{drop} < (P_{max,Rx} \cdot K)/(P_{amp})$ and $R_{drop} > (P_{min,Rx} \cdot K \cdot N)/P_{amp}$ in an $N$ channel system. This assumes that for each $K$ a dedicated 1x$K$ splitter is used, and no additional attenuation is present. Assuming $P_{amp}$= 20 dBm, a 96-channel system, where 8 channels should be dropped, and a dynamic range between $P_{Rx,max}$ = +3 dBm and $P_{Rx,min}$ = -23 dBm for the Rx, the ratio should be 3.8 %<$R_{drop}$<16 %. For the add ratio, $R_{add}$, holds that the input power of each added channel needs to be equal to the channels passing through the wavelength blocker, i.e.

$$\frac{P_{Tx}}{K} \cdot R_{add} = \frac{P_{amp}(1-R_{drop})}{a_{WB} \cdot N}(1-R_{add}), \text{ and therefore}$$

$$R_{add} = \frac{P_{amp}(1-R_{drop})}{\frac{P_{Tx}}{k}a_{WB}N+P_{amp}(1-R_{drop})}. \quad (1)$$

Here, $P_{Tx}$ is the output power of the transmitter (Tx), which can be in ideal case be adjusted in a range of some dB and $a_{WB}$ is the insertion loss of the WB. For the example above, and $P_{Tx}$ = 3 dBm, $a_{WB}$ =12 (8 dB raw loss of WB +4 dB for compensation of +-2 dB of channel power variations), the minimum drop ratio leads to an add ratio of $R_{add}$ = 20 % and the maximum drop ratio leads to an add ratio of 17.9 %. In this case the loss of the node with minimum add ratio would be either 13.2 dB ($R_{drop}$ = 3,8 %) or 13.7 dB ($R_{drop}$=16 %). A higher add ratio would increase the loss. For a fibre loss of 0.25 dB/km and a variable gain of the EDFA of up to 30 dB, the maximum span length would be 65.2 km ($R_{drop}$=16 %), which lies well in the span range of a metro network.

Another important parameter in the system is the OSNR. In some state-of-the-art coherent transponders, additional optical amplifiers are integrated at the Tx. Besides increasing $P_{Tx,max}$, they also add broadband noise to the signal if they do not use a tuneable amplified spontaneous emission (ASE) filter after the amplifier. If, e.g. the OSNR of the Tx is 40 dB, after coupling 96 channels with the same Tx, the OSNR is then reduced to 20.2 dB. The OSNR reduction along the transmission line can be calculated from the power levels and the noise figure of the EDFAs (see, e.g. [10], chapter 4.2.3), i.e. the OSNR after the $M$th span would be

$$OSNR_{M,EDFA} = \frac{OSNR_{TX,N}P_{in,EDFA}}{P_{in,EDFA}+Mhf_cB_{ref}F_NOSNR_{TX,N}}. \quad (2)$$

Here, $OSNR_{Tx,N}$ is the OSNR per channel after coupling together $N$ channels, $P_{in,EDFA}$ the input power to the EDFA after the span, $h$ Planck's constant, $f_C$ the frequency of the signal carrier and $B_{ref}$, the reference bandwidth in Hz for measuring the OSNR, per definition 0.1 nm in wavelength, which corresponds to ~ 12.5 GHz. $F_n$ is the noise figure of the EDFA. If in each node unfiltered channels are added, additional noise overlaps the noise from the transmission. If $K$ channels are added, the $OSNR_{add,K}$ for each channel would be the OSNR of the transmitter itself divided by the numbers of added channels. E.g. if the OSNR of the Tx is 40 dB, after coupling together 8 channels, this would be reduced to 31 dB. The OSNR after adding channels can be calculated by

$$OSNR_{Add} = \frac{OSNR_{through}OSNR_{add,K}}{OSNR_{through}+OSNR_{add,K}}, \quad (3)$$

With $OSNR_{through}$ are the channels which are passing through the node. Now, combining this, the OSNR after the $M$th node with $K$ channels added in each node would be

$$OSNR_{M,E+A} = \frac{P_{in,EDFA}}{Mhf_cB_{ref}F_N+P_{in,EDFA}\left(\frac{M}{OSNR_{add,K}}+\frac{1}{OSNR_{TX,N}}\right)}. \quad (4)$$

Since the strongest OSNR degradation occurs at the MCEN, one might think of using a multiplexer filter there and only have a filterless coupling in the AMENs.

All the calculations and assumptions here were made for fixed power and gain values with variations. In further investigations, gain ripples and power variations of up to 3 dB should be taken into account.

## 3. Experimental evaluation

The experimental setup corresponded to the structure shown in Fig. 2. Two commercially available coherent transponders, capable to terminate 200 Gbit/s DP-16QAM or 100 Gbit/s DP-QPSK signals were used as channels under test (CUTs). Forward error correction was included, leading to a symbol rate of ~ 34 GBaud. The pre-FEC BER was calculated from the corrected errors. Although the FEC for the transponders here was higher than 2e-2, we considered 2e-2 as a FEC limit, since this commonly used for soft-decision FEC. The 88 interferer channels (loaders) were shaped out of an ASE noise source, using two EDFAs and a subsequent Finisar waveshaper as in [4]. The spectrum ranged from 191.55 THz to 196 THz, and the CUTs are put at the 5[th] and the 85[th] channel slot (191.75 THz and 195.8 THz respectively) to determine differences due to EDFA noise tilt. All transmit channels were added at the MCEN terminal node on the left. Between the spans, couplers and WBs with $R_{drop}$ = 20 % and $R_{add}$ = 20 %, as shown in Fig. 2, were inserted. The WBs had a loss of 12 dB. According to Eq. (1), in this setup, $R_{add}$=20 % would be optimum if 9 channels are added and $P_{TX,max}$= 3 dBm. The investigated topology consisted of 4×40 km standard single-mode fiber spans (SSMF). The 40-km spans had a loss of 10 dB each. Fig. 3 shows the spectrum of all channels after the first span and after the forth span. It can be seen that, although the power is equalized for all spans, the EDFA noise shows a tilt, leading to a tilted OSNR as well. Longer wavelengths (and thus smaller frequencies, e.g. channel 5) have a higher OSNR than shorter wavelengths (e.g. channel 85).

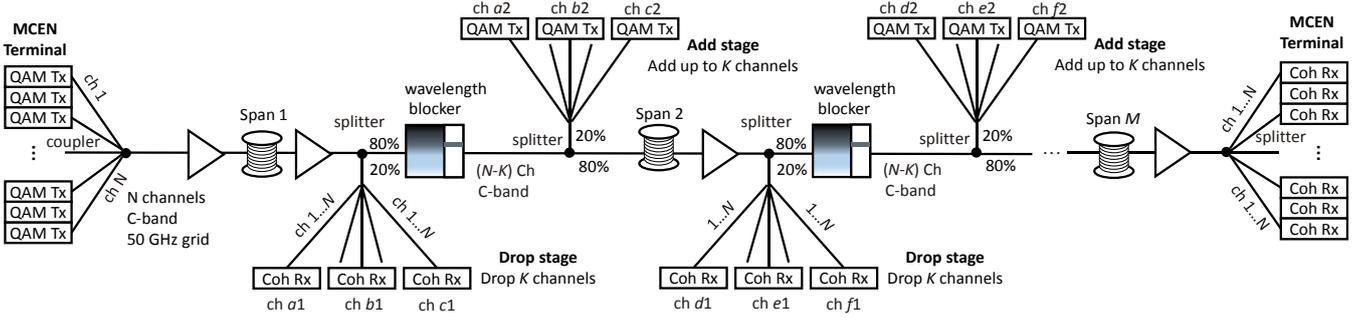

Fig. 2. Structure of the semi-filterless network concept with wavelength blockers. Up to *N* channels can be transmitted, at each Add/Drop stage *K* channels are dropped and added (QAM Tx: quadrature amplitude modulation transmitter, Coh Rx: coherent receiver).

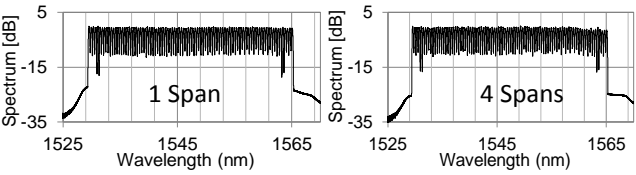

Fig. 3 Spectrum of the 90 channels after 1 span and after 4 spans of 40 km.

Since in this setup $R_{add}$ and $R_{drop}$ are fixed, the experiment focussed on the influence of transmit OSNR. The OSNR $OSNR_{TX,N}$ after combining all channels was varied. If a pure filterless coupling is used, the OSNR of one channel would be 19.5 dB higher than this value (90 channels ~ 19.5 dB). The OSNR and the bit error ratio (BER) for each CUT were determined at each drop stage and after the last EDFA. The loss from the drop coupler to the Rx was 13 dB, which is equivalent to dropping up to 20 channels. The total input power to the Rx was 2.5 dBm, i.e. -17 dBm per channel.

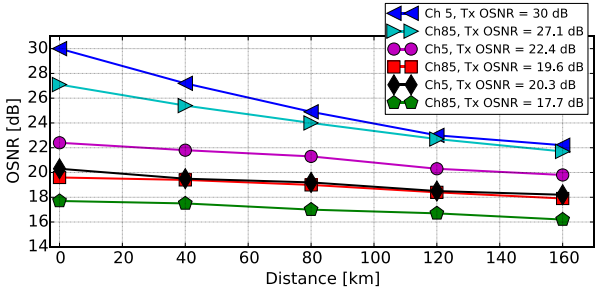

Fig. 4 OSNR vs. distance for different Tx OSNRs.

Fig. 4 shows the evolution of the OSNR over the 4 spans. The 40-km spans degrade the OSNR significantly only if the Tx OSNR is high – for values less than 20 dB the decrease is quite slow. The values for $OSNR_{TX,N}$ = 30dB at channel 5 and $OSNR_{TX,N}$ = 27.1dB at channel 85 were measured simultaneously as were the values for 22.4 dB and 19.6 dB, and 20.3 dB and 17.7 dB, respectively The BER values corresponding to the OSNRs shown in Fig. 4 are plotted in Fig. 5 for DP-QPSK and in Fig. 6 for DP-16QAM. It can be seen that the Tx OSNR is a critical parameter, and 22 dB Tx OSNR would be needed to bridge all spans with full capacity, if the BER should stay below 2e-2, which is a commonly used limit for soft decision FEC. Considering a 90 channel terminal in which each transmitter adds broadband noise, this requires an OSNR of at least 42 to 45 dB at the transmitter, if a noise tilt of 3 dB is present. However, referring to (4), also the noise from added channels needs to be taken into account, which was not done during the experiment.

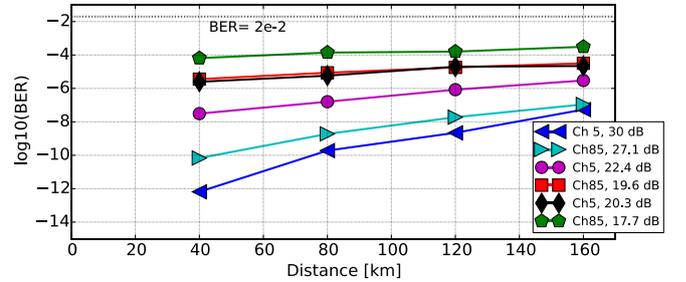

Fig. 5 BER vs. distance for DP-QPSK and OSNR of Fig. 4.

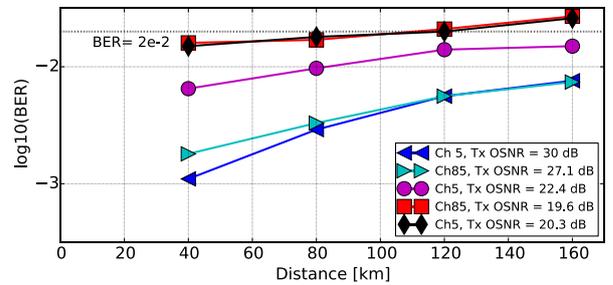

Fig. 6 BER vs. distance for DP-16QAM and OSNR measured in Fig.4. 30/27.1 dB and 22.4/19.6 dB are measured simultaneously, for 20.3 dB only channel 5 could be detected.

## 4. Conclusion

A semi-filterless metro network architecture based on wavelength blockers was presented and evaluated. Our investigation shows that the add and drop ratios for the splitters have to be chosen carefully, dependent on system parameters. Wavelength blockers enhance the transmission capacity at lower cost compared to wavelength selective switches and enable channel equalization. The transmitter OSNR is a critical parameter if no ASE filters are used in the add paths.

## 5. Acknowledgement

This work has received funding from the European Union's Horizon 2020 research and innovation programme under grant agreement No. 761727 (METRO-HAUL).

## 6. References

[1] https://metro-haul.eu accessed 13 April 2019

[2] L. E. Nelson, et al., "Colorless reception of a single 100Gb/s channel from 80 coincident channels via an intradyne coherent receiver," IEEE Photonics Conference 2012, Burlingame, CA, 2012, pp. 240-241. doi: 10.1109/IPCon.2012.6358581.

[3] C. Tremblay, et al., "Filterless Optical Networks: A unique and Novel Passive WAN Network Solution", Proc. 12 th Optoelectronic and Communications Conference / 16 th Intl. Conference on Integrated Optics & Optical Fiber Communication IEICE , pp. 466-467, 2007.

[4] R. Emmerich, R. Elschner, C. Schmidt-Langhorst, et al., "Colorless C-band WDM system enabled by coherent reception of 56-GBd PDM-16QAM using an high-bandwidth ICR with TIAs," 2017 Optical Fiber Communications Conference and Exhibition (OFC), Los Angeles, CA, 2017, paper M2C.3, pp. 1-3, doi: https://doi.org/10.1364/OFC.2017.M2C.3

[5] G. Serafino et al., "Semi Filter-Less Drop & Waste Network Demonstration with Integrated SOI Optical Filter," 2017 European Conference on Optical Communication (ECOC), Gothenburg, 2017, pp. 1-3. doi: 10.1109/ECOC.2017.8346157

[6] O. Ayoub, et al., "Filterless and Semi-Filterless Solutions in a Metro-HAUL Network Architecture," 2018 20th International Conference on Transparent Optical Networks (ICTON), Bucharest, Romania, 2018, pp. 1-4. doi: 10.1109/ICTON.2018.8473651

[7] H. Miyata et al., "Fully dynamic and reconfigurable optical add/drop multiplexer on 0.8 nm channel spacing using AOTF and 32-wave tunable LD module," Optical Fiber Communication Conference. Technical Digest Postconference Edition. Trends in Optics and Photonics Vol.37 (IEEE Cat. No. 00CH37079), Baltimore, MD, USA, 2000, pp. 287-289 vol.4. doi: 10.1109/OFC.2000.869486

[8] Y. Sakurai et al., "LCOS-Based Wavelength Blocker Array With Channel-by-Channel Variable Center Wavelength and Bandwidth," in IEEE Photonics Technology Letters, vol. 23, no. 14, pp. 989-991, July15, 2011. doi: 10.1109/LPT.2011.2148702

[9] F. Paolucci et al., "Filterless Optical WDM Metro Networks Exploiting C+L Band," 2018 European Conference on Optical Communication (ECOC), Rome, 2018, pp. 1-3. doi: 10.1109/ECOC.2018.8535529

[10] M. Eiselt, K. Grobe, "Wavelength Division Multiplexing: A Practical Engineering Guide", Wiley, 1st Edition, 2013.